\documentclass{aa}
\usepackage{psfig}
\newcommand{\FeII}{Fe\,{\sc ii}}
\newcommand{\FeX}{Fe\,{\sc x}}
\newcommand{\OIII}{O\,{\sc iii}}
\begin{document}

   \thesaurus{03     
               11.01.2;  
               11.19.1;  
               11.17.3;  
               13.25.2} 

   \title{Identification of a complete sample of northern ROSAT All-sky
         Survey X-ray sources}

   \subtitle{VII. The AGN subsample}

   \author{
           I. Appenzeller \inst{1,4}
          \and
          F.-J. Zickgraf \inst{2}
          \and J. Krautter \inst{1}
          \and W. Voges \inst{3}}

   \offprints{I. Appenzeller}

   \institute{Landessternwarte Heidelberg,
              K\"{o}nigstuhl, D-69117 Heidelberg, Germany
         \and Hamburger Sternwarte, Gojenbergsweg 112,
              D-21029 Hamburg, Germany
         \and Max-Planck-Institut f\"{u}r extraterrestrische Physik,
              Gie\ss enbachstra\ss e, D-85740 Garching, Germany 
         \and Max-Planck-Institut f\"{u}r Astronomie,
              K\"{o}nigstuhl, D-69117 Heidelberg, Germany
             }



   \date{Received; accepted}

   \maketitle

   \begin{abstract}
   In order to derive statistical properties of a complete X-ray selected 
   sample of AGN we used the classification spectra of the RASS Selected 
   Areas-North survey to study the luminosities, redshifts , X-ray/visual
   flux ratios, line widths, and various correlations between observed 
   parameters of all Seyfert galaxies and QSOs in this sample. On the basis 
   of these data we discuss implications for the current AGN models.   
 
   \end{abstract}


\section{Introduction}
In two earlier papers (Zickgraf et al. 1997a, Appenzeller et al. 1998,
hereafter referred to as ``Paper I'' and ``the Catalog'', respectively) we 
described and presented the optical identification of a complete sample of 
X-ray sources discovered in the ROSAT All-sky Survey (RASS, Voges et al. 
1999). A statistical analysis of these identifications (which in the 
literature are also referred to as the ``RASS Selected Area-North survey'')
has been presented by Krautter et al. (1999). Subsamples of the Catalog have 
been discussed by by Zickgraf et al. (1998b), Mujica et al. (1999), 
and Appenzeller et al. (2000). The present paper presents a discussion of 
the Seyfert galaxies and QSOs listed in the Catalog as main optical 
counterparts of RASS X-ray sources with an Identification Quality Index $Q = 1$
or $Q = 2$ (i.e. reliable identification on the basis of our spectra and/or
on the basis of reliable literature data). Since the Catalog lists no 
Seyfert galaxy and only one QSO with $Q = 3$ (``uncertain identification'') 
the present paper covers practically all objects of these types in our 
survey.

Not included in the following discussion are the BL Lac objects (which are 
discussed elsewhere), the two LINERs, and the 7 AGN for which due the 
weakness of the lines or inadequate $S/N$ no subclass could be assigned in
the Catalog. Also omitted are 3 Seyfert galaxies and one QSO which
- although inside the error circle of an X-ray position - were not regarded 
to be the main source of the observed X-ray radiation since another (in
our opinion dominant or more likely optical counterpart was (also) 
present at the position in question.      

As described in Paper I our identifications and classifications were
based on low-resolution spectra obtained  with a multi-object spectrometer 
at a 2.1-m telescope. These spectra were exposed to reach a $S/N$ suitable 
for a reliable classification and redshift derivation. Originally there 
were no plans for a quantitative evaluation of these spectra. However, 
after photometric calibration the great majority of the spectra turned out 
to be of sufficient quality to measure the basic properties of the 
AGN spectra, such as line strengths and line widths. Therefore, we made 
use of the opportunity provided by these data to derive statistical
information on a well defined complete sample of X-ray selected AGN.

Throughout this paper Seyfert 1 galaxies and QSOs will be discussed jointly
since - according to our classification criteria - these two classes 
are distinguished only by being below or above a certain (arbitrarily chosen)
luminosity level. Included in this group are all AGN with permitted 
emission lines (or components of the permitted emission lines) measurably
broader than the forbidden lines and/or showing \FeII\ emission. Hence
Seyfert 1.5-1.9 and NLS1 galaxies (as defined by V\'eron-Cetty and V\'eron,
2000, following Osterbrock and Pogge, 1985) are part of our Seyfert 1 and 
QSO sample. The (relatively few) Seyfert 2s in the Catalog are discussed 
separately in Section 3.

In order to keep the classification criteria uniform for the whole sample,
the [\OIII ]5007\AA /H$\beta $ flux ratio (which is measurable in only part
of our spectra) was not used as a classification criterion to discriminate
between Seyfert 1s and 2s (Osterbrock and Pogge, 1985, Laor, 2000). 
However, in all spectra containing the [\OIII ]/H$\beta $ region,
the corresponding flux ratio was determined and compared to the critical
value 3. While all galaxies classified as Seyfert 1s from their 
line profiles showed (as expected)
[\OIII ]/H$\beta <3$, we found (as described in Section 3) 6 objects 
classified as Seyfert 2s with [\OIII ]/H$\beta <3$. Because of this 
small number the classification of these objects has 
no influence on the 
statistical conclusions derived for the Seyfert 1s and QSOs. However,
as discussed in Section 3,  the much smaller Seyfert 2 sample is
affected if the [\OIII ]/H$\beta $ flux ratio is used as classification
criterion.     

Not discussed in the present paper are AGN number counts, $\log N - \log S$ 
distributions, and the implications of our data for the unresolved X-ray 
background, since the corresponding results have already been presented by 
Krautter et al. (1999) and Miyaji et al. (2000). Instead, the present 
paper will concentrate on conclusions concerning the AGN physics.     

\section{The QSOs and Seyfert 1 galaxies}
\subsection{General properties}
According to Krautter et al. (1999) 228 sources in our Catalog (about
34\% of the total) were identified as Seyfert 1s or QSOs. Apart from stars
(37\%) these objects form the largest subgroup in our sample. In the 
Catalog we list the X-ray flux $f_{\rm X}$, the visual magnitude $V$, the 
X-ray/visual index $I_{\rm XV}$  and the redshift $z$. From these 
data we calculated approximate absolute visual magnitudes $M_{V}$ 
(assuming $h = 0.75$, $q_{0}$ = 0.5, $\Lambda$ = 0 and negligible interstellar
extinction). The distribution of the resulting $M_{V}$ values is plotted 
in Fig. 1. The redshift distribution of our Seyfert 1s and QSOs is given 
in Fig. 2. As shown by the figures, our sample is dominated by 
moderate-luminosity objects ($M_{V} \approx -23\pm 3$) with an average
redshift $<z>$ = 0.40. Only 8\% of our Seyfert 1s and QSOs have 
redshifts $>$ 1.0. Only one QSO (RX\,J1028.6-0844, $z$ = 4.28, cf. Zickgraf 
et al., 1997b) has a redshift $>$ 2.2. The shape of the distributions 
in Figs. 1 and 2 indicate that our sample provides an essentially complete
inventory for the local ($z \le 0.2$) AGN population while for higher 
redshifts only the progressively rarer objects with high (X-ray)
luminosity are detected by the RASS.
        
\begin{figure}
\hspace{-.5cm}
\psfig{figure=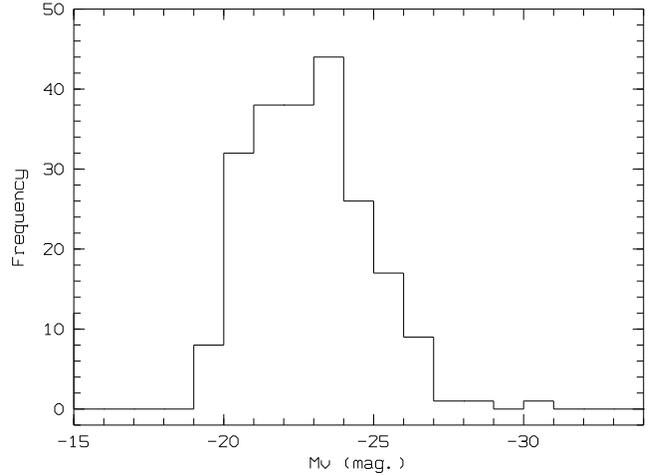,width=9.5cm,angle=-90}
\caption{Visual luminosity distribution of the Seyfert 1s and QSOs}
\end{figure}

In Fig. 3 we present the distribution of the X-ray-visual index $I_{\rm XV}
= \log (f_{\rm X}/f_{V})$, which provides a measure for the relative
strength of the X-ray and visual emission of an object. For AGN
$I_{\rm XV}$ is approximately linearly related to the $\alpha _{\rm ox}$ 
index, which is used for the same purpose in part of the literature
(see e.g. Stocke et al. 1991). For our ROSAT data we have with good
approximation $\alpha _{\rm ox} = 1.44 - 0.39$$I_{\rm XV}$. The accuracy
this relation for our data can be estimated from Fig. 4.    
Our $I_{\rm XV}$ values are of the same order as those found in other AGN surveys 
(e.g. Stocke et al., 1991).

Since our sample is X-ray flux-limited, we have to expect a selection effect
in the sense that apparently faint objects with low $I_{\rm XV}$ will not 
be included in our survey and only very high $I_{\rm XV}$ objects 
are observed at the faint end. As shown in Fig. 5, this selection effect
is clearly present. However, as shown by Fig. 6,  no such effect is  
conspicuous in our $I_{\rm XV}$-redshift relation, since (apart from the 
very bright objects, which are all at small redshifts) there exists no 
strong correlation between
redshift and apparent brightness in our sample. 
Therefore, the observed $I_{\rm XV}$ distribution 
is probably characteristic for the RASS AGN independent of 
the redshift. (The presence
of 3 low-redshift objects with $I_{\rm XV} < -1.0$ in Fig. 6 is probably caused by 
an overestimate of the luminosity of these relatively faint AGN 
due to a contamination of the photometry by their host galaxy light.
Hence, these 3 $I_{\rm XV}$ values are probably lower limits only).

\begin{figure}
\hspace{-.5cm}
\psfig{figure=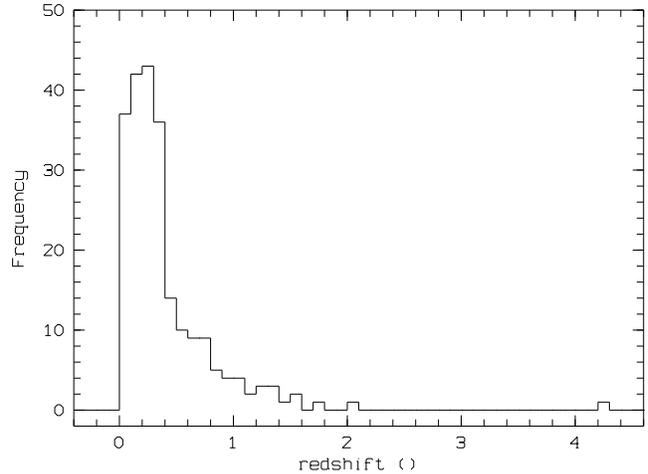,width=9.5cm,angle=-90}
\caption{Redshift distribution of the Seyfert 1s and QSOs}
\end{figure}

The characteristic property of Seyfert 1s and QSOs is the presence
of broad (BLR) emission lines . While the instrumental resolution
(corresponding to about 750 kms$^{-1}$) did not allow us to resolve the 
forbidden line profiles, the BLR profiles were usually well resolved 
and intrinsic line widths $>$ 500 kms$^{-1}$ could normally detected 
from the broadened profiles. In the Standard Model of AGN the widths
of the BLR lines are assumed to be caused by the motions of the line emitting 
plasma in the potential of the central black hole. Since the distance of the
BLRs to the central continuum source can (in principle) be estimated
using reverberation techniques, the line widths and their distributions 
provide important information on the central masses and mass distributions.
A direct comparison of all BLR line widths in our sample is complicated by 
the redshift range of our spectra. In order to allow a direct and unbiased 
comparison, we, therefore, had to restrict our analysis to a spectral region
which is common to at least most of our spectra. Best suited for this 
purpose turned out to be the region of the H$\beta $ line.

\begin{figure}
\hspace{-.5cm}
\psfig{figure=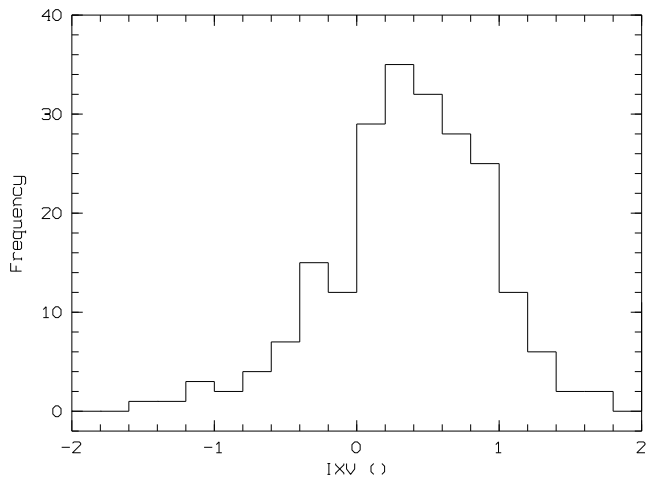,width=9.5cm,angle=-90}
\caption{Distribution of the X-ray-visual index $I_{\rm XV}$ for the
Seyfert 1s and QSOs}
\end{figure}

Depending on the position of the object in the field and the $S/N$ of the 
spectra this line falls into our observed spectral range for
redshifts of about 0 $<$ $z$ $<$ 0.8. For all objects in this range with 
spectra of of adequate $S/N$ we measured the FWHM and FWZI line widths
of H$\beta $. Since the quality of our spectra is not sufficient to allow a 
reliable decomposition of the profiles into various components, the FWHM
measurements refer to the full line profiles, including broad and narrow
components. (The FWZI widths measure the BLR components only, but, due to
difficulties defining the continuum level, FWZI values are normally less 
reliable). The resulting FWHM distribution is plotted in Fig. 7. (The FWZI
distribution is broader by about a factor of 2 but qualitatively similar).

The distribution in Fig. 7 shows a relatively large 
fraction (18\% $\pm $ 3\%) of objects with FWHM $<$ 2000 km$^{-1}$.
However, a comparison with the literature indicates that this fraction
is not unusual for AGN. Stephens (1989) finds for a small ($N = 42$) 
X-ray selected sample 24\%. In the optically selected sample
of Boroson and Green (1992) the corresponding fraction is 23\% $\pm $ 5\%,
which agrees within the error limits well with our X-ray selected sample.
If the 6 narrow-line objects with [\OIII ]/H$\beta < 3$ mentioned above
are added to our Seyfert 1 sample, our FWHM $<$ 2000 kms$^{-1}$ fraction 
increases to 21\% $\pm $ 3\%, providing an even better agreement with
Boroson and Green. The fact that Puchnarewicz et al. (1992) find in a
sample of 17 Seyfert 1s with ultra-soft X-ray spectra 9 objects 
(53 \%) with
H$\beta $ FWHM $<$ 2000 kms$^{-1}$ may indicate a relation with the 
X-ray spectral index. However, because of the limited X-ray spectral 
information for our objects (see below) this relation cannot be tested 
with our data.    
Most of the objects with 
FWHM $<$ 2000 km$^{-1}$ also have BLR components with larger FWHM. Only three
objects classified as Seyfert 1s in our Catalog are NLS1 galaxies without
detectable Balmer line components of FWHM $>$ 2000 km$^{-1}$, but with 
strong \FeII\ emission. The presence of many objects with strong 
narrow Balmer components and weak broad components argues for a smooth
transition between the NLS1s and other Seyfert types. (The fact that 
Engels and Keil (2000) 
in their analysis of a different sample of X-ray selected
AGN find a higher fraction of NLS1s is probably due to different
classification or selection criteria).        
        
\begin{figure}
\psfig{figure=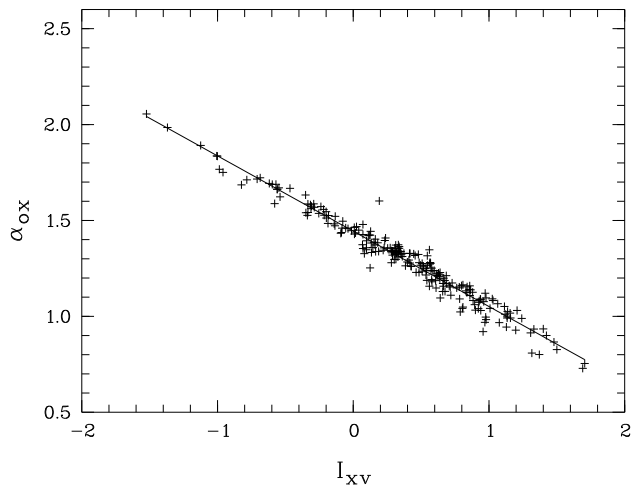,width=8.5cm,angle=-90,bbllx=45pt,bblly=30pt,bburx=600pt,bbury=700pt,clip=}
\caption{The relation of the indices $\alpha _{\rm ox}$ and $I_{\rm XV}$ for the 
Seyfert 1s and QSOs in our sample. The solid line corresponds to the
linear function given in the text.}
\end{figure}

As shown by Fig. 7, the line widths cover a range exceeding a factor 
of 10. The broadest H$\beta $ line was observed for the object 
RX\,J1021.6-0327 = Akn 241 (FWHM = 9600 km$^{-1}$, FWZI = 17 900 km$^{-1}$).
The spectrum of this object seems to show some other spectral 
peculiarities, which have to be studied with better $S/N$ and higher
resolution, however. 

According to the AGN Standard Model the line width distribution of the 
broad lines can be caused (a) by variations of the depth of the
gravitational potential of the line forming region or (b) by variations 
of the orientation of the rotation axis relative to the line of sight
to the observer. For disklike rotating emission regions with uniform
velocities the theory predicts for (b) a distribution with a minimum 
at low velocities and a maximum and cutoff at the high velocity limit.
The distribution in Fig. 7 is obviously very different, indicating that the 
line widths variations are probably dominated by intrinsic orbital
velocity differences (i.e. variations of the potential) of the BLRs.

\begin{figure}
\hspace{-.5cm}
\psfig{figure=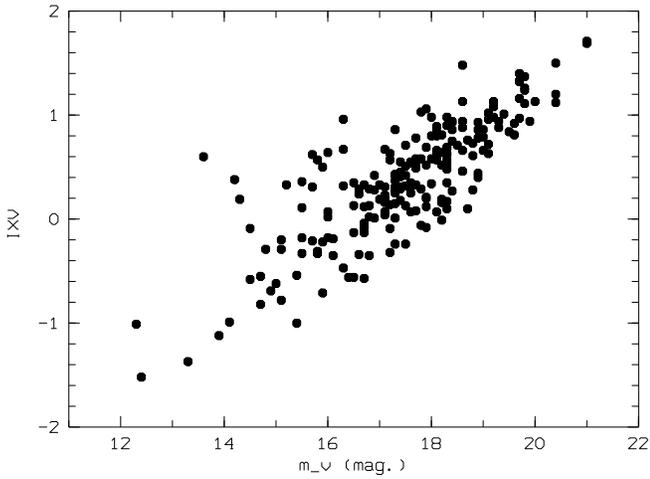,width=9.5cm,angle=-90}
\caption{The X-ray-visual index $I_{\rm XV}$ of the Seyfert 1s and QSOs
as a function of the apparent visual magnitude $m_{V}$}
\end{figure}

Except for the hydrogen and helium lines \FeII\ multiplets are normally
the most conspicuous emission features in the visual spectra of the
Seyfert 1s and QSOs. Their strength is normally measured by the \FeII\ 
index (flux ratio) $R4570$ = Fe 4570\AA /H$\beta $. Unfortunately
our spectra were normally not of sufficient quality to measure the   
\FeII\ 4570\AA \ blend directly. On the other hand, for 63 objects it was 
possible to derive the total strength of the \FeII\ (37,38) blends near
4570\AA \ and the \FeII\ (48,49) blend near 5300\AA . Assuming that the 
relative strength of the \FeII\ multiplets is constant, we converted these 
measurements to approximate $R4570$ values using the well observed 
(strong-\FeII ) Seyfert 1 galaxy I Zw I (Phillips 1977, Boroson and Green
1992) for calibration. As in other Seyfert 1 samples the great majority
(80\%) of our $R4570$ values fall into the interval 0.1 - 1.0. Our mean
$R4570$ value 0.7 is somewhat higher than the normally quoted
average for Seyfert 1s (0.4, Osterbrock 1977, Bergeron and Kunth, 1984),
although this difference is not significant in view of our approximate
method and and the size of our sample. Nevertheless, the fact that 
the average \FeII\ emission is certainly not lower in our X-ray
selected sample than in normal Seyfert 1 galaxies seems to argue against
the result of Lawrence et al. 1997, who (on the basis of a smaller sample)
find the \FeII\ emission to be anticorrelated to the X-ray emission 
in Seyferts. As pointed out below, we also found no anticorrelation 
(or correlation) between $R4570$ and and $I_{\rm XV}$ for our sample.
        
\begin{figure}
\hspace{-.5cm}
\psfig{figure=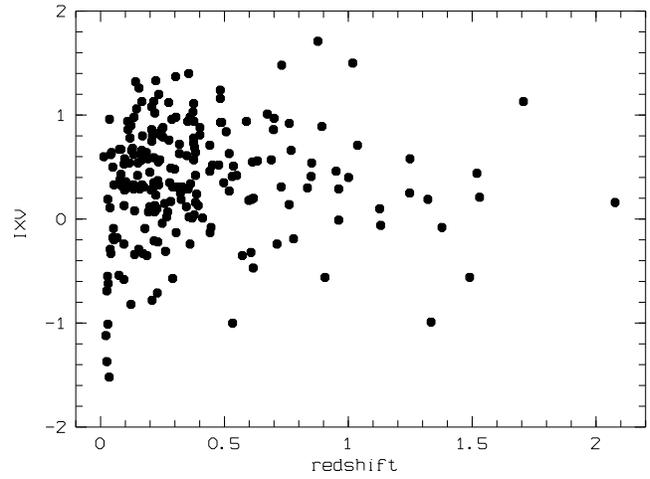,width=9.5cm,angle=-90}
\caption{The X-ray-visual index $I_{\rm XV}$ of the Seyfert 1s and QSOs
as a function of the redshift $z$}
\end{figure}

One object in our sample (RX\,J0757.0+5832) shows (as already noted in the 
Catalog) exceptionally strong \FeII\ emission ($R4570 \approx $ 2.05).   
 
\subsection{Correlations}
As pointed out e.g. by Dahari and De Robertis (1988) there are few strong 
correlations between different AGN properties. Hence it was no surprise
that we found (apart from trivial relations, such as between FWHM and 
FWZI) few correlations in our data. In particular we find no significant 
correlation between the absolute visual brightness and the H$\beta $
line width, although such a correlation seems to be present in other
AGN samples (e.g. Miller et al. 1992). Our Fig. 8, showing the
observed relation, present essentially a random scatter apart from the fact 
that the few objects with H$\beta $ FWHM $>$ 8000 km$^{-1}$ all have 
luminosities below $M_{V} = -22$, while the most luminous QSOs show
moderate line widths. Within the Unified AGN Model this could perhaps be 
explained assuming that the high-FWHM objects are likely seen edge-on 
with the central light source partially obscured by a circumnuclear dust 
torus. However, in this case we may expect to find redder than average 
$B-V$ values and different $I_{\rm XV}$ values for the high line width
objects. Since this is not observed, we conclude that the high line width of
low-luminosity objects in Fig. 8 is probably not caused by an inclination
effect.

In Fig. 9 we plotted the \FeII\ emission strength (expressed in $R4570$) as a 
function of the H$\beta $ FWHM line widths for all those objects where 
both these quantities could be measured. Our plot confirms the well
known anticorrelation between \FeII\ emission and BLR line widths for
Seyfert 1s and QSOs (see e.g. Zheng and Keel 1991, Wang et al. 1996,
Lawrence et al. 1997). On the other hand, in contrast to
Lawrence et al. (1997), we found in our sample 
no indication of any correlation between the \FeII\ emission strength 
and the X-ray loudness (as expressed by $I_{\rm XV}$).

\begin{figure}
\hspace{-.5cm}
\psfig{figure=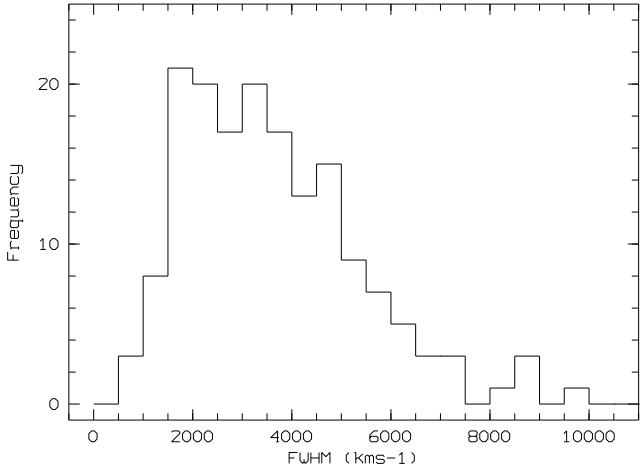,width=9.5cm,angle=-90}
\caption{Distribution of the H$\beta $ FWHM line widths of the Seyfert 1s 
and QSOs}
\end{figure}

As pointed out e.g. by Mushotzky et al. (1993), luminous AGN normally
tend to show lower $I_{\rm XV}$ (or steeper $\alpha _{\rm ox}$) values than 
low-luminosity objects. As shown in Fig. 10 this correlation is also
indicated in our data. However, apart from the large scatter in our data,
the relation derived here is probably affected (i.e. weakened) by the 
selection effect demonstrated in Fig. 4.  

Boller et al. (1996), Laor et al. (1997) Grupe et al. (1999), and others
pointed out a correlation between the ROSAT spectral index and the 
H$\beta $ line width for Seyfert galaxies and QSOs. Our relatively 
large sample of Seyfert galaxies with ROSAT X-ray data provide in principle 
a possibility to study this correlation. Therefore, we calculated the ROSAT 
photon index $\Gamma $ for all our objects. Unfortunately for most of our 
objects the photon counts turned out to be much too low to derive $\Gamma $
with an acceptable accuracy. For only 13 objects with good H$\beta $ data
we were able to determine $\Gamma $ with a mean error $<$ 1.0. These data, 
plotted in Fig. 11, are consistent with the known anticorrelation between 
the BLR line widths and $\Gamma $.

\begin{figure}
\hspace{-.5cm}
\psfig{figure=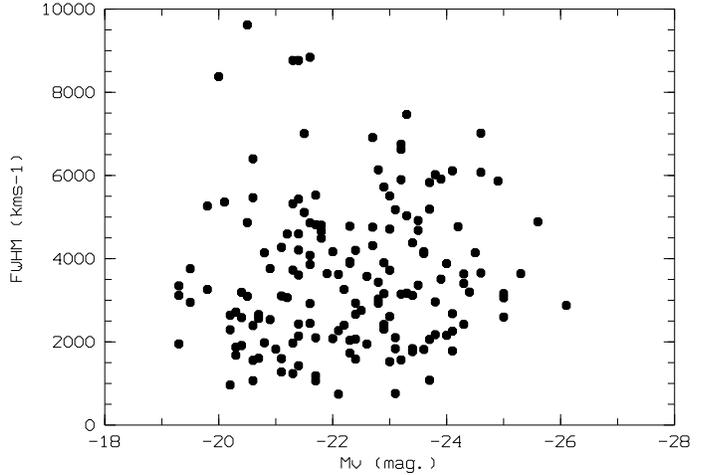,width=9.5cm,angle=-90}
\caption{H$\beta $ FWHM line widths
as a function of the  visual luminosity}
\end{figure}
        
\begin{figure}
\hspace{-.5cm}
\psfig{figure=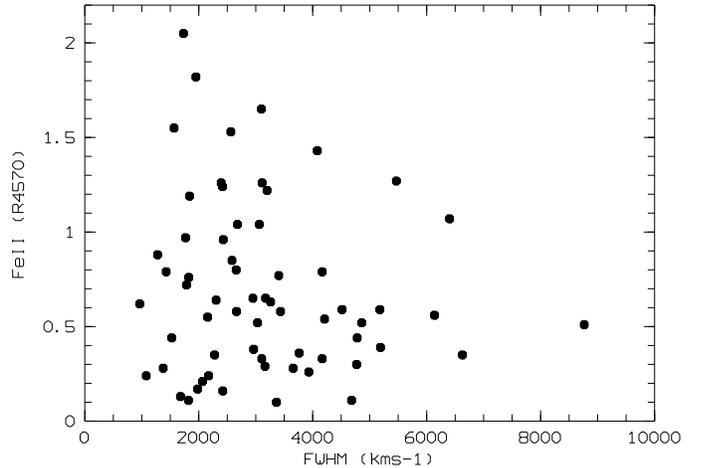,width=9.5cm,angle=-90}
\caption{\FeII\ emission strength ($R4570$) 
as a function of the H$\beta $ FWHM line widths}
\end{figure}

Most of our spectra are not of sufficient quality to detect the weak 
forbidden high ionization lines (FHILs) or ``coronal'' lines of the Seyfert
spectra. In only 5 objects in our sample the [\FeX ] lines were strong 
enough to be visible on our spectra. Interestingly, two of these 5 objects
(RX\,J0707.2+6435 and RX\,J1218.4+2948) are also among the 3 objects with 
$\Gamma > 3.0$, supporting the existence of a correlation between $\Gamma $ 
and the FHIL strength, as proposed by Erkens et al. (1997).

\begin{figure}
\hspace{-.5cm}
\psfig{figure=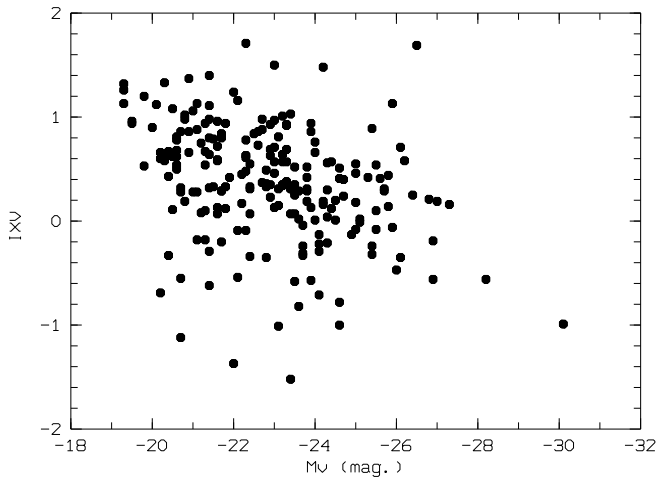,width=9.5cm,angle=-90}
\caption{The X-ray-visual index $I_{\rm XV}$ of the Seyfert 1s and QSOs
as a function of the visual luminosity}
\end{figure}

\section{The Seyfert 2 galaxies}
While the Catalog lists 228 Seyfert 1s and QSOs, only 16 Seyfert 2s were
identified as a primary optical counterpart of one of the observed 
X-ray sources. Moreover, our classification criteria for Seyfert 2s 
(Seyfert spectrum without detectable broad lines or line components
and without detectable \FeII\ emission) could have resulted in a 
contamination of the Seyfert 2 sample by Seyfert 1.5-1.9 and NLS1 objects
with weak BLR components and weak \FeII\ emission (since broad line 
components and weak \FeII\ blends are more strongly affected by a low $S/N$
than narrow spectral features covering fewer pixels). Moreover,
while (as pointed out in Section 1) the Seyfert 1 sample is practically 
unaffected by the definitions used, a Seyfert 1 classification based
entirely on the [\OIII ]/H$\beta $ flux ratio may result in a different
Seyfert 2 sample. In order to estimate
these potential effects we re-inspected the spectra of all objects
classified as Seyfert 2s in our Catalog and derived 
[\OIII ]/H$\beta $ flux ratios and approximate limits for
the presence of undetected BLR components and \FeII\ blends. In this process 
we found one case (RX\,J2218.6+0802) which definitely should have been 
classified differently. New measurements of the corresponding spectrum
resulted in FWHM(H$\beta $) = 1520 $\pm $ 120 kms$^{-1}$, 
FWHM([\OIII ]) = 878 $\pm $ 200 kms$^{-1}$ (both uncorrected for the 
instrumental profile)  and [\OIII ]/H$\beta $ = 2.0. Obviously, 
this object should have been classified as a NLS1. Five additional 
objects do not show detectable BLR or \FeII\ features but definitely 
[\OIII ]/H$\beta $ $<$ 3.0. In a classification based on the
[\OIII ]/H$\beta $ ratio alone, these objects would also have
to be classified as NLS1s. On the other hand, at least 7 of the objects 
listed in our Catalog as Seyfert 2s have [\OIII ]/H$\beta $ $>$ 3.0, Balmer 
BLR contributions $<$ 10\% and \FeII\ $R4570 <$ 0.2. Hence, we conclude that
at least about 50\% of the optical counterparts classified as Seyfert 2s 
in our Catalog are bona fide type 2 objects, independently of the exact 
classification criteria and a possible contamination with type 1 objects due 
to the inadequate $S/N$ of some of our spectra.

\begin{figure}
\hspace{-.5cm}
\psfig{figure=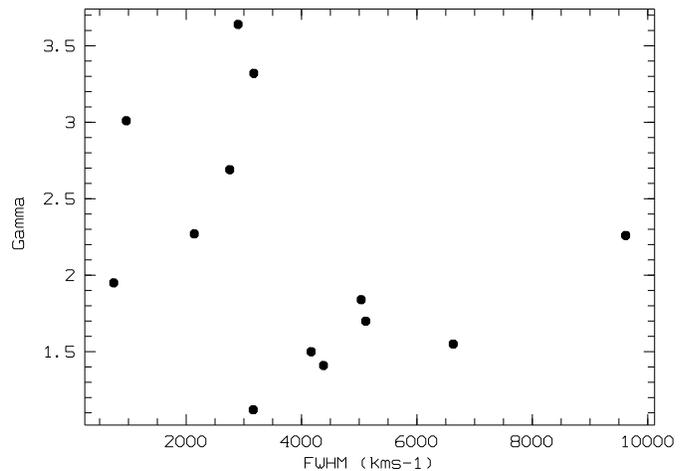,width=9.5cm,angle=-90}
\caption{The ROSAT photon index $\Gamma $
as a function of the H$\beta $ FWHM line widths}
\end{figure}

In view of the uncertainties and ambiguities of the Seyfert 2 sample,
in the following all statistical results are listed separately for the
full original sample of the Catalog (16 objects) followed [in brackets]
by the corresponding value for the minimal sample of 7 reliable objects.
Because of the small size of the samples, the discussion will
be restricted to statistical means.

Compared to the Seyfert 1 and QSO sample the most obvious difference is 
the significantly lower mean luminosity (and correspondingly lower mean 
redshift)
of the Seyfert 2s (reflecting also the scarcity of narrow line QSOs).
The visual luminosities of our Seyfert 2s covered the range 18.1 $<$ 
$-M_{V}$ $<$ 22.8 [18.1 $<$ $-M_{V}$ $<$ 22.1] with a median value of
$-20.5$, as compared to $-22.9$ for the Seyfert 1 and QSO sample.
Hence the observed Seyfert 2s have on average only about 1/10 of the 
visual luminosity of the observed Seyfert 1s and QSOs. The Seyfert 2 
redshifts are all below 0.5 with an average of 0.18 [.22]. On the other 
hand, the $I_{\rm XV}$ distribution of the Seyfert 2s (median value 
$<I_{\rm XV}> = 0.53 \pm .15 [.7 \pm .3]$ is not significantly different 
from that of the Seyfert 1s and QSOs ($<I_{\rm XV}> = 0.40 \pm .04$).

For our flux limited sample the (visual and X-ray) Seyfert 2 to 1
luminosity ratio of 1/10 means that the volume in which we observe the
Seyfert 2s is only about 3\% of that of the Seyfert 1s and QSOs.
Since the relative fraction of the Seyfert 2s in our Catalog
([3 \%] to 7 \%) is rather close to this number
the Seyfert 2s detected in the RASS, although on average 
much fainter, are found to have about the same (or slightly larger)
space density than the observed Seyfert 1s. Equal space densities
of Seyfert 1s and Seyfert 2s have also been estimated for the general
AGN population (Simkin et al., 1980, V\'eron and V\'eron-Cetty, 1986).
In this respect our X-ray selected AGN obviously show the same 
behavior as the general AGN population.

\section{Implications for the AGN models}
As shown in Fig. 1 the luminosity distribution of the X-ray selected AGN 
has a FWHM of about 5 mag., corresponding to a factor 100. On the other 
hand Fig. 3 shows a FWHM of the X-ray/visual flux ratio of only a factor 
of $\approx $ 10. This seems to indicate that the basic physical 
conditions of the X-ray and visual continuum emitting volumes of the AGN
are similar and are not strongly dependent on the luminosity. Moreover,
the fact that the X-ray selected Seyfert 1s and 2s have similar 
mean $I_{\rm XV}$ values supports the assumption of a similar physical structure
of all Seyfert nuclei and QSOs.

Assuming that the H$\beta $ FWZI corresponds to about twice the maximum 
orbital velocity of the BLR plasma, we obtain for our sample
velocities of 1000 $< v_{\rm orbit} < 9000$ kms$^{-1}$. If these values 
correspond to circular orbits in the potential of a black hole, the 
observed velocity range can be converted into a radial distance 
range for the BLRs of $6 \times 10^{2} < r/R_{\rm s} < 5 \times 10^{4}$
(where $R_{\rm s} = 2{\rm G}M/{\rm c}^{2}$ is the BH's Schwarzschild radius). 
The observed particularly high BLR line widths for some 
low-luminosity Seyfert 1s and the absence of such high velocities 
for the more luminous QSOs (Fig. 8) could in this case be explained 
by the fact that the intense radiation field of the luminous QSOs
prevents the existence of relatively cool Balmer line emitting 
BLR gas at low $r/R_{\rm s}$ values, while at low luminosities the BLR
gas can exist in a large range of distances. Less clear is the 
interpretation of the NLS1 in this scheme. (For a recent compilation
of proposed explanations see Komossa and Janek, 2000). While the 
assumption of a low central mass being the origin of the absence of
broad line components (Boller et al., 1999; Laor, 2000) seems plausible,
other explanations cannot be ruled out on the basis of the present data.

\begin{acknowledgements}
     We thank Drs. Thomas Boller, Max Camenzind, Stefan Wagner 
     and Ari Laor for
     critically reading the manuscript and for valuable comments.
\end{acknowledgements}

\end{document}